%% file: main.tex
\documentclass{article}
\usepackage{spconf,amsmath,graphicx,hyperref}
\usepackage{kotex}

\usepackage{xspace}
\usepackage{xcolor}
\usepackage[most]{tcolorbox}
\usepackage{booktabs}
\usepackage{makecell}
\usepackage{amsmath}
\usepackage{tabularx}
\usepackage{pifont} 
\usepackage{amsfonts}
\usepackage{multirow}
\usepackage{graphicx}
\usepackage[capitalise]{cleveref}
\usepackage{subcaption}

\newcommand{\ours}{\textbf{\texttt{TruS}}\xspace}
\newcommand{\eg}{\textit{e.g.}}
\newcommand{\ie}{\textit{i.e.}}

\definecolor{red}{rgb}{0.8, 0.2, 0.2}
\definecolor{darkred}{rgb}{0.6, 0.1, 0.05}
\definecolor{blueish}{rgb}{0.0, 0.3, .6}
\newcommand{\bp}{\textbf{P}}
\newcommand{\forget}{\text{Opt}}
\newcommand{\retain}{\text{Ret}}

\newcommand{\tparagraph}[1]{\noindent\textbf{#1}}
\newcommand{\cmark}{\textcolor{blueish}{\ding{51}}}%
\newcommand{\xmark}{\textcolor{red}{\ding{55}}}%
\title{Erasing Your Voice Before It’s Heard: \\ Training-free Speaker Unlearning for Zero-shot Text-to-Speech}
\name{Myungjin Lee$^*$ \quad\quad Eunji Shin$^*$ \quad\quad Jiyoung Lee$^\dagger$\thanks{$^*$Equal contribution. \quad $^\dagger$Corresponding author.}}
\address{School of AI and Software, Ewha Womans University}

\begin{document}
\ninept
\maketitle

\input{tex/0_abstract}

\input{tex/1_intro}
\input{tex/3_method}
\input{tex/4_experiments}
\input{tex/5_conclusion}

\clearpage
\bibliographystyle{IEEEbib_short}
\bibliography{strings,refs}

\end{document}

%% file: tex/0_abstract.tex
\begin{abstract}
Modern zero-shot text-to-speech (TTS) models offer unprecedented expressivity but also pose serious crime risks, as they can synthesize voices of individuals who never consented.
In this context, speaker unlearning aims to prevent the generation of specific
speaker identities upon request.
Existing approaches, reliant on retraining, are costly and limited to speakers seen in the training set.
We present \ours, a training-free speaker unlearning framework that shifts the paradigm from data deletion to inference-time control.
\ours steers identity-specific hidden activations to suppress target speakers while preserving other attributes (\eg, prosody and emotion).
Experimental results show that TruS effectively prevents voice generation on both seen and unseen opt-out speakers, establishing a scalable safeguard for speech synthesis.
The demo and code are available on \url{http://mmai.ewha.ac.kr/trus}.
\end{abstract}


%% file: tex/1_intro.tex
\section{Introduction}\label{sec:intro}

The advancement of zero-shot TTS systems recently reached a level of expressivity and naturalness that makes it attractive for a wide range of applications in accessibility~\cite{szekely2025voice} and content creation~\cite{jia2025broadcast}.  
Meanwhile, their ability to generalize across speakers~\cite{le2023voicebox, eskimez2024e2,chen2024f5} introduces a profound risk: such models may generate the voices of real individuals who never consented to their use~\cite{ alali2025partial, tomashenko2025first}.
Unlike traditional text or image generation, speech carries strong biometric identity cues, making unauthorized voice synthesis particularly sensitive from both privacy and security perspectives.

Existing countermeasures fall short of addressing this problem at its root.
Watermarking techniques~\cite{juvela2024collaborative,zong2025audiomarknet} are inherently post hoc: they can only verify or trace synthetic speech after it has been generated, offering no protection against real-time misuse.
Voice anonymization (VA)~\cite{srivastava2020design,panariello2024voiceprivacy} aims to protect privacy by transforming speech to replace the original identity, typically through speaker disentanglement and re-synthesis.
It follows a substitution technique, replacing one identity with another, whereas our objective is to prohibit a specific identity from being generated at all.
A more extreme solution would be to prohibit voice generation altogether. 
Yet this would undermine the utility of TTS technology, which is critical for accessibility and human-computer interaction.

On the one hand, machine unlearning has emerged to remove the influence of specific training data from learned models. 
However, most existing methods, including gradient-based~\cite{bourtoule2021machine, baumhauer2022machine, warnecke2023machine}, knowledge distillation-based~\cite{chundawat2023can,fan2024salun}, and latent manipulation approaches~\cite{guide}, have primarily been developed in the field of image or text generation.
These approaches often involve retraining of the model, leading to high computational cost, unstable convergence, and unintended forgetting. 
In TTS, only a few attempts have been made~\cite{kim2025not}:
sample-guided unlearning (SGU) and teacher-guided unlearning (TGU).
While TGU adapts distillation to forget training-set speaker identities, it still requires substantial retraining whenever new unlearning requests arise, making it impractical for scalable deployment.
In addition, they cannot handle unseen individuals, even though real-world opt-out requests are most likely to come from users outside the training set.
Taken together, these limitations reveal a deeper structural gap in current approaches to speaker protection in TTS.
Existing methods are either reactive (\eg, watermarking), substitutive (\eg, VA), or training-bound (\eg, retraining-based unlearning), and thus fail to provide a practical mechanism for enforcing user intent at the time of generation.

\begin{figure}
    \centering    \includegraphics[width=.9\linewidth]{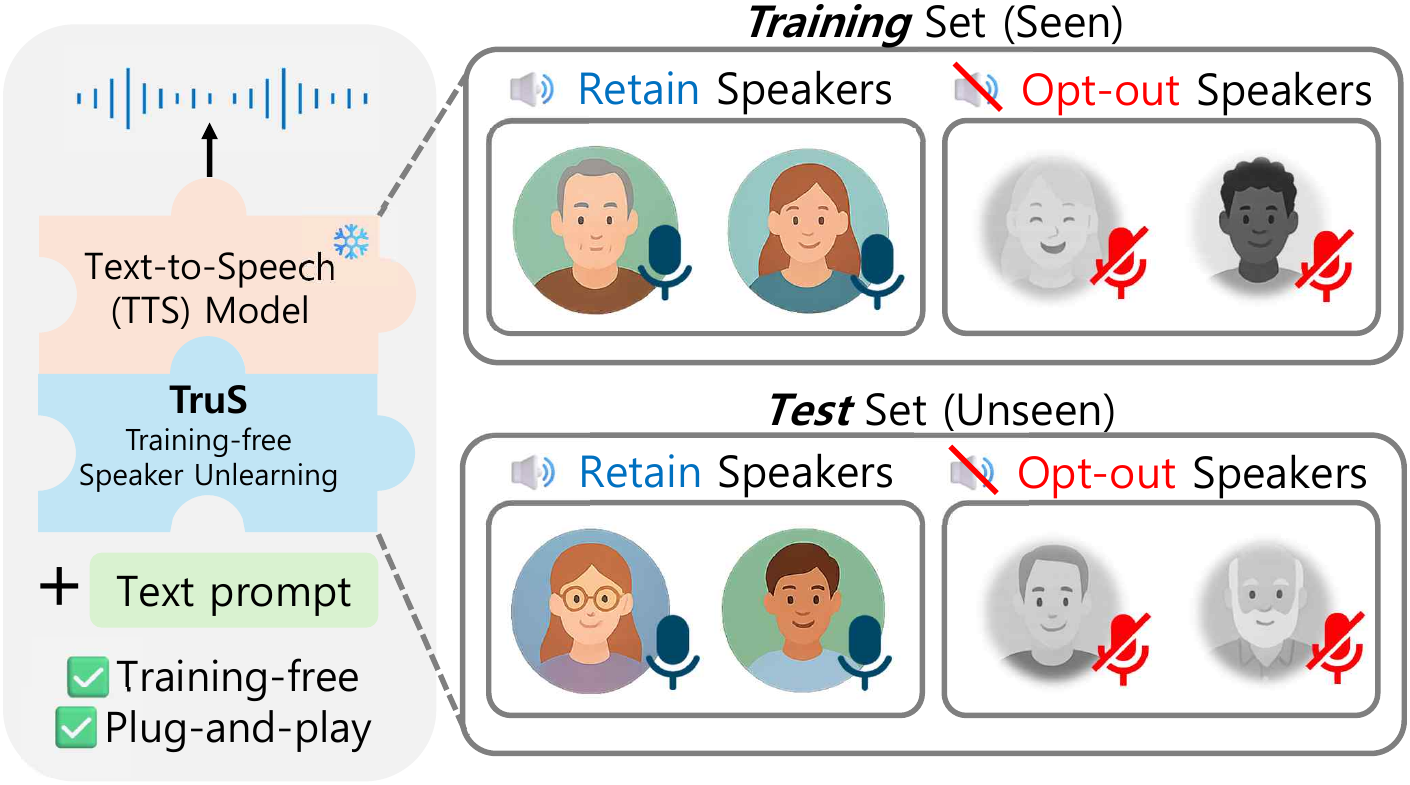}
    \vspace{-1em}
    \caption{Illustration of training-free speaker unlearning.}
    \label{fig:teaser}\vspace{-1em}
\end{figure}

To address this gap, we propose \ours, the first training-free, inference-time framework for opt-out speaker unlearning in zero-shot TTS, which allows individuals to explicitly request that their voices not be synthesized.
\ours directly intervenes in the TTS model's internal activations to suppress identity-specific information.
This reframes unlearning as a user-driven safeguard, enabling immediate and sequential handling of unlearning requests in speech generation.
Notably, \ours generalizes beyond the seen opt-out speakers in the training set and is capable of blocking the synthesis of unseen opt-out speakers, as depicted in \cref{fig:teaser}.

\ours is motivated by the observation that speaker identity is encoded in structured directions within the hidden representations of modern TTS models.
Even though EmoSteer~\cite{xie2025emosteer} has attempted to modulate emotional prosody by heuristically selecting top-k activation channels in TTS, they apply the same fixed rule across inputs to remove any dynamic adaptability.
In contrast, our goal is to prevent the synthesis of specific speaker identities while preserving essential paralinguistic attributes. 
For a small set of retain speakers' utterances, \ours first generates an identity prototypical embedding (ID-prototype) with intermediate features of TTS models.
We statistically analyze the similarity according to identity in each layer to automatically select steering blocks.
\ours dynamically guides hidden representations with such an ID-prototype to solely revise the target (\ie, opt-out) speaker's identity.
Experimental results demonstrate that \ours effectively suppresses target speakers' identities without retraining, achieving comparable unlearning performance to existing methods.
Crucially, \ours is the first to generalize to unseen opt-out speakers, extending unlearning beyond the training set to block the generation of voices that mimic individuals.

\begin{figure}[t]
    \centering
    \includegraphics[width=\linewidth]{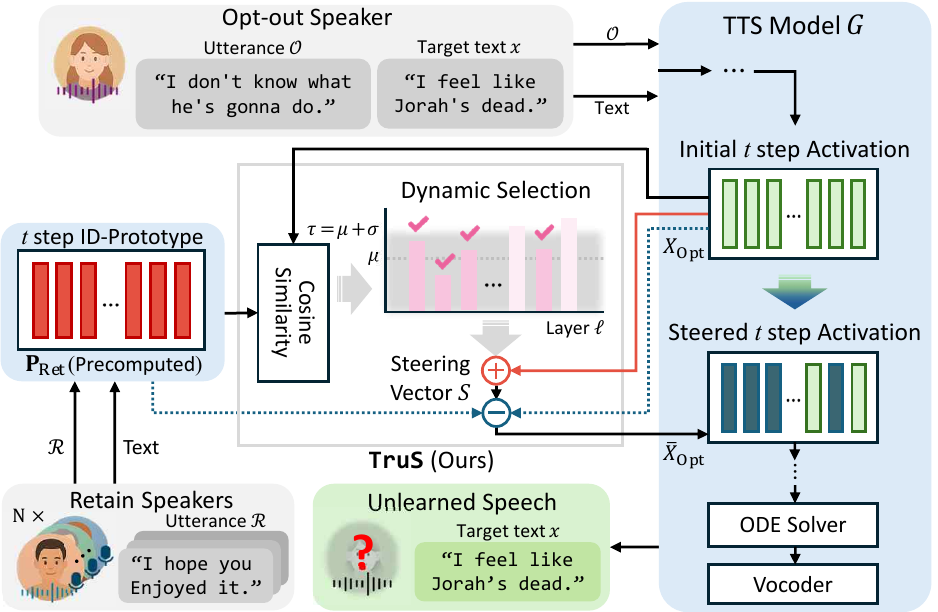}
    \vspace{-1em}
    \caption{The overall framework of \ours, working with TTS models at inference time.
    Feature activations at layers and generation steps are optionally steered based on the dynamically selective threshold.
    With only a single utterance example of a target who requests to opt out, our method controls to suppress the identity-related activations without additional training.}
    \label{fig:method}
    \vspace{-.5em}
\end{figure}

In summary, our contributions are three-fold:\vspace{-3pt}
\begin{itemize}
    \item We propose a novel framework, dubbed \ours, the first training-free unlearning for zero-shot TTS that constrains models from generating both seen and unseen speaker identities in speech generation.\vspace{-4pt}
    \item We design a dynamic steering mechanism to selectively control identity-related activations with ID-prototype.\vspace{-4pt}
    \item \ours demonstrates comparable performance to tremendously trained baselines, and it further supports opt-out and sequential unlearning requests in a scalable manner.\vspace{-5pt}
\end{itemize}

%% file: tex/3_method.tex
\section{Method}\vspace{-.5em}
\subsection{Motivation and problem formulation}\vspace{-.5em}
Previous unlearning in TTS, such as TGU and SGU~\cite{kim2025not}, rely on retraining the model on a filtered dataset with the target speaker data removed.
However, as the generalized capability of TTS models is increased~\cite{du2024cosyvoice2,ozspeech2025}, the models impose a well-formed speaker embedding space.
Consequently, we speculate that removing specific speaker data during training does not guarantee the complete elimination of information.
Furthermore, TGU~\cite{kim2025not} requires additional training whenever a new user requests opt-out.
Motivated by prior works~\cite{rimsky2024steering, turner2023steering} in natural language processing, we introduce that steering internal representations in TTS models is an effective and efficient solution to control the speaker identity.

\input{tex/fig_3}

Our new system, \ours, enables opt-out unlearning by maintaining a query pool of speaker embeddings for individuals who request their voice not to be synthesized, defined as opt-out sets for training and test data, respectively.
Let $\mathcal{R}$ denote the set of reference utterances from retain speakers, and let $\mathcal{O}$ denote the set of reference utterances from opt-out speakers, \ie, speakers who explicitly request that their voices not be synthesized.
Let $u \in \{\mathcal{R} \cup \mathcal{O} \}$ denote a reference utterance.
Given a TTS model $G(x, u)$ that generates speech from text $x$ conditioned on a reference utterance $u$, our objective is to control the generated speech based on whether $u$ belongs to $\mathcal{O}$.
Our \ours is built upon the recent fabulous TTS model, F5-TTS~\cite{chen2024f5}.
In the following, we describe our approach on top of this baseline for clarity, though our method is generally applicable to intermediate blocks of other DiT-based~\cite{dit} TTS architectures.
\vspace{-5pt}

\subsection{Identity-specific steering vector}\vspace{-.5em}
To suppress the identity-related feature of the opt-out speaker in the generated speech, \ours steers hidden activations with dynamic selection of salient layers.
Existing works~\cite{chensteering,touvron2023llama,turner2023steering} for steering in LLMs often require a lot of hyperparameters to control the tradeoff between the fidelity to the prompt and the generation quality.
Our \ours overcomes this challenge through dynamic steering with only one-shot reference example.

Specifically, we prebuild a prototypical identity vector, shortly \textit{ID-prototype}, at each DiT block $\ell \in \{1, \dots, L \}$ in timestep using $N$ utterances from retain speakers in $\mathcal{R}$.
Note that each utterance sample is from a different speaker.
In particular, the outputs of FFN in DiT blocks contain strong timbre and identity signals after non-linear channel mixing~\cite{lin2025ffn}. 
This design allows us to capture identity-specific differences without degrading intelligibility or prosody.
The extracted hidden activations $X^{(\ell, t)}_{\retain}$ over the retain speakers are averaged to build an ID-prototype, denoted as $\bp^{(\ell, t)}_{\retain}$ at $\ell$-th block in flow step $t \in \{T, \dots, 1\}$, which serves as our base centroid point:\vspace{-.7em}
\begin{equation}
    \bp^{(\ell, t)}_{\retain}= \frac{1}{N}{\sum_{n=1}^{N}~ X^{(\ell, t)}_{\retain(n)}} 
\end{equation}

During inference of TTS models, our \ours simultaneously computes steering vectors and controls the hidden activations to conceal the target speaker's identity in generated speech.
Given an utterance of an opt-out speaker in $\mathcal{O}$, the corresponding target activation $X^{(\ell, t)}_\forget$ is taken from the FFN outputs at each $\ell$-th DiT block.
Based on the prebuilt ID-prototype $\bp^{(\ell, t)}_{\retain}$, our identity-specific steering vector $S^{(\ell, t)}$ for the target speaker is defined as $L_2$-normalized of the difference of those activations at the $\ell$-th block in flow step $t$:
\begin{equation}
    S^{(\ell, t)} \;=\; \frac{X^{(\ell, t)}_{\forget} - \bp^{(\ell, t)}_{\retain}}
{\left\| X^{(\ell, t)}_{\forget} - \bp^{(\ell, t)}_{\retain} \right\|_2}
\end{equation}
Therefore, $S^{(\ell, t)}$ represents the step-wise identity-related direction of the target speaker within the latent space.
They form the basis for suppressing identity-related representation while preserving linguistic content and paralinguistic attributes.

\input{tex/tab_main}
\vspace{-5pt}
\subsection{Dynamic selection of steering layers}\vspace{-.5em}
Even though the identity-specific cues are encoded to build the basis of activation steering, we argue that \textit{`not all layers contribute equally to maintain speaker identity'}.
As illustrated in \cref{fig:step_cos}, cosine similarities between $\bp^{(\ell, t)}_{\retain}$ and $X^{(\ell, t)}_\forget$ are dynamically varying at each flow step $t$ in the generation process.
With such a key finding, layers that exhibit lower cosine similarity are considered potential intervention points, since they diverge more strongly between the target speaker and ID-prototype.

Specifically, given a target sample that requests opt-out, cosine similarity $c^{(\ell,t)}$ is measured between ID-prototype $\bp^{(\ell,t)}_{\retain}$ and the corresponding target activation $X^{(\ell,t)}_{\forget}$ at all flow steps and layers.
Lower similarity indicates a larger deviation from the ID-prototype, suggesting identity-specific activations of the target speaker at the corresponding layer and step.
To automate layer selection, we propose a dynamic ID threshold $\tau$ based on global layer-wise statistics.
For each layer $\ell$, we first aggregate step-wise similarities
\begin{equation}
    \bar{c}^{(\ell)}= \frac{1}{T} \sum_{t=1}^{T} c^{(\ell,t)}.
\end{equation}
Using the set of layer-wise similarities $\{\bar{c}^{(\ell)}\}$, the global mean $\mu$ and variance $\sigma^2$ are computed over all layers using the set of layer-wise mean similarities $\bar{c}^{(\ell)}$:
\begin{equation}
    \mu = \frac{1}{L}\sum_{\ell=1}^{L} \bar{c}^{(\ell)},
    \qquad
    \sigma^2 = \frac{1}{L}\sum_{\ell=1}^{L}
    \bigl(\bar{c}^{(\ell)}-\mu \bigr)^2.
    \label{eq:layer_stats} 
\end{equation} 
Finally, the dynamic threshold is defined as
\begin{equation}
    \tau = \mu + k\,\sigma,
\end{equation}
where $k$ adjusts the tolerance band to balance a trade-off between identity suppression and preservation of speech quality (\ie, larger $k$ selects fewer layers and smaller $k$ conservative selection). 
Layers whose average similarity is below this threshold are collected as a set of intervention layers.
As a result, multiple layers may be selected for a given opt-out speaker, with the specific set determined by the layer-wise similarity distribution.
Since the statistics are computed per target sample, the threshold $\tau$ naturally varies across different opt-out speakers.

Given the selected set of intervention layers, we further introduce a finer step-level filtering within each layer.
For each selected layer $\ell'$, intervention is restricted to flow steps whose similarity satisfying $c^{(\ell',t')} < \bar{c}^{(\ell')}$. 
This two-stage criterion induces a sparse set of layer-step intervention points.
By dynamically tailoring both layer and step selection per sample, steering avoids excessive or misplaced interventions while maintaining generation ability.

\subsection{Unlearning via inference-time steering}\vspace{-.5em}
The hidden activations are modified on-the-fly during the denoising process. 
For each chosen layer $\ell'$ and flow step $t'$, the pre-computed steering vector $S^{(\ell',t')}$ is applied to suppress identity-specific information.
In order to remove only the component aligned with the identity-related direction while preserving other linguistic and prosodic content, the projection of the activation onto the steering vector is subtracted:
\begin{equation}
    \bar{X}^{(\ell', t')}_{\forget}={X}^{(\ell', t')}_{\forget}- \alpha ~(~{X}^{(\ell', t')}_{\forget} \cdot {S}^{(\ell', t')}~){~S}^{(\ell', t')},
    \label{eq6}
\end{equation}
where $\alpha$ is a steering strength to control the intervention intensity.
Without a burden of training, \ours therefore suppresses the contribution of the target speaker to intermediate representations at inference time with the additional benefit of successfully preserving the naturalness of speech.\vspace{-.5em}

%% file: tex/fig_3.tex
\begin{figure*}
    \centering
    \begin{subfigure}[b]{.32\linewidth}
        \centering
        \includegraphics[width=\linewidth]{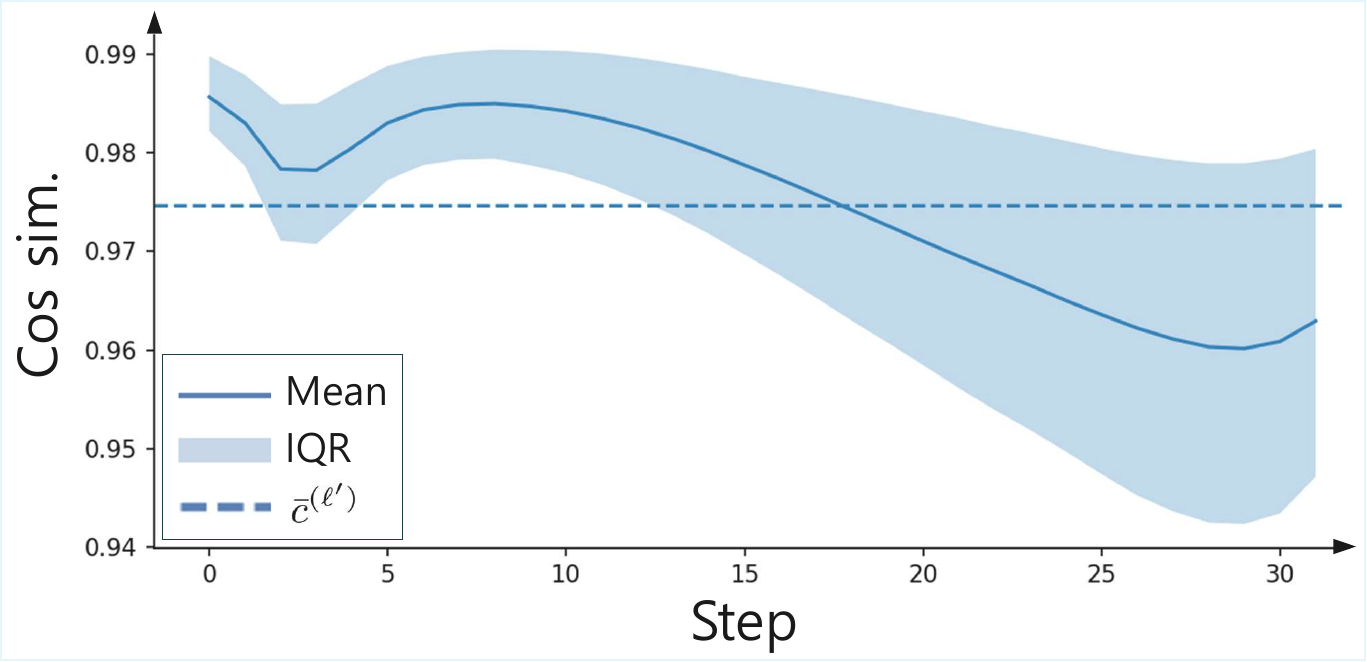}\hfill\vspace{-0.5em}
        \caption{1st block layer}
        \label{fig:3a}
    \end{subfigure}
    \begin{subfigure}[b]{.32\linewidth}
        \centering
        \includegraphics[width=\linewidth]{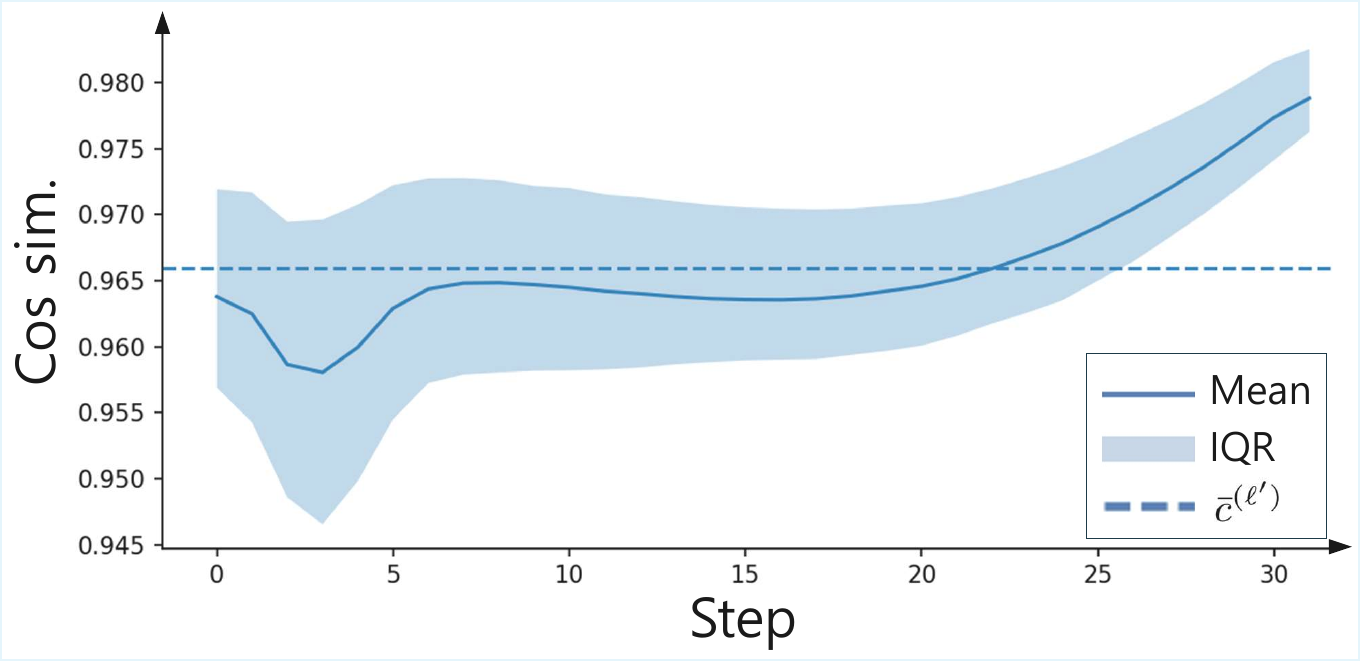}\hfill\vspace{-0.5em}
        \caption{12th block layer}
        \label{fig:3b}
    \end{subfigure}
    \begin{subfigure}[b]{.32\linewidth}
        \centering
        \includegraphics[width=\linewidth]{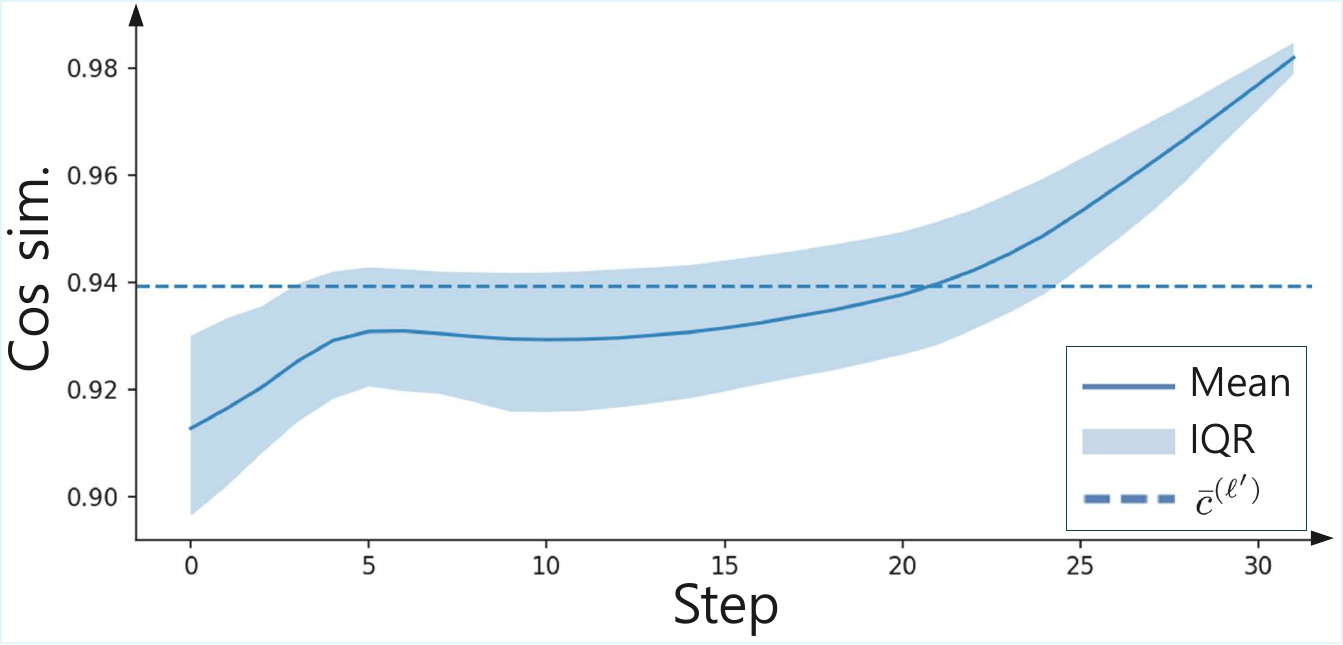}\hfill\vspace{-0.5em}
        \caption{20th block layer}
        \label{fig:3c}
    \end{subfigure}
    \vspace{-.7em}
    \caption{Examples of step-wise cosine similarities between hidden activations of target speaker and ID-prototype, at 1st, 12th, 20th layers. Similarity increases in later steps at later layer, and decreases in earlier ones, indicating the need for dynamic layer- and step-specific steering.}\vspace{-.5em}
    \label{fig:step_cos}
\end{figure*}

%% file: tex/tab_main.tex
\begin{table*}[!t]
    \centering
    \small
    \resizebox{0.8\linewidth}{!}{
    \begin{tabular}{l c ccc ccc}
        \toprule
        Methods & Training hours &  WER-R$\downarrow$ & SIM-R$\uparrow$ & WER-SO$\downarrow$ & SIM-SO$\downarrow$ & Spk-ZRF-R & Spk-ZRF-SO$\uparrow$ \\
        \midrule
        F5-TTS~\cite{chen2024f5}           & - & 1.95 & 0.678 & 3.36 & 0.657 & 0.908 & 0.925 \\
        F5-TTS-FT~\cite{chen2024f5}                      & 52 & 2.07 & 0.654 & 3.13 & 0.656 & 0.911 & 0.924 \\
        SGU~\cite{kim2025not}   & 48 & 2.12 & 0.290 & 3.70 & 0.106 & 0.935 & 0.959 \\
        TGU~\cite{kim2025not}   & 430 & 2.21  & 0.549 & 4.03 & 0.510 & 0.921 & 0.933 \\
        \midrule
        \ours                & 0 &  1.95$^\dagger$ & 0.678$^\dagger$ & 3.25  & 0.477 & 0.908$^\dagger$ &  0.929 \\
        \bottomrule
    \end{tabular}
    }
    \vspace{-.5em}
    \caption{Quantitative results on LibriSpeech (-R) and the seen opt-out set (-SO) on Emilia. `F5-TTS-FT' denotes the finetuned F5-TTS only with the retain set of Emilia. Since our method solely works for the seen opt-out set, $^\dagger$ scores follow the original model.
    Training hours are reported based on two A6000 GPUs.}\label{tab:table1}
\end{table*}

%% file: tex/4_experiments.tex
\section{Experiments}\vspace{-.7em}
\subsection{Experimental settings}\vspace{-.5em}
\tparagraph{Datasets.}
Our baseline is F5-TTS~\cite{chen2024f5} pretrained on Emilia~\cite{he2024emilia}, a large-scale multi-lingual corpus.
We note that only the English subset is considered in this study.
For the standard unlearning evaluation protocol, 10 training-set speakers are designated as the \textit{seen opt-out} (SO) set, with 300 seconds held out for test. 
The other speakers in Emilia constitute the \textit{retain} set. 
For zero-shot evaluation, LibriSpeech test-clean corpus~\cite{panayotov2015librispeech} is employed to evaluate the generalized unlearning capability. 
The \textit{unseen opt-out} (UO) set of LibriSpeech consists of 10 sexually-balanced speakers ($\sim$300 seconds each), and the other speakers belong to the retain set.
This protocol enables systematic evaluation of whether the method suppresses opt-out speakers while avoiding unintended degradation for unseen voices to the retain speakers.
Finally, emotional fidelity is assessed on CREMA-D~\cite{cremad}, where 10 speakers are selected as an unseen opt-out set, and 30 utterances per speaker are generated.

\tparagraph{Implementation details.}
Our method entirely works at inference time, operating on a pretrained open-source TTS model, F5-TTS~\cite{chen2024f5} without any additional finetuning. 
For a fair comparison, SGU~\cite{kim2025not} and TGU~\cite{kim2025not} are reimplemented based on F5-TTS with the authors' released code. 
In practice, we empirically set the steering strength to $\alpha=1.2$ to balance forgetting effectiveness and generation quality.

\tparagraph{Metrics.}
Following prior work~\cite{kim2025not}, we assess the performance with SIM~\cite{le2023voicebox}, WER, Spk-ZRF~\cite{kim2025not}, and SIM-Emo.
SIM measures the identity similarities between generated and reference speech, using features computed by a pretrained speaker verification model~\cite{Desplanques_2020}. 
WER quantifies linguistic fidelity by comparing the transcription of generated speech with the reference text,  where transcriptions are obtained using a pretrained Whisper large-V3~\cite{whisper}.
Spk-ZRF~\cite{kim2025not} measures the degree of randomness in the generated voices for opt-out speakers. 
Finally, we employ emotion2vec~\cite{ma2024emotion2vec} to measure SIM-Emo, the emotional similarities between the generated and reference speech.
We report all metrics under three evaluation conditions: retain (R), seen opt-out in Emilia (SO), and unseen opt-out in LibriSpeech or CREMA-D (UO).

\input{tex/tab_libri}
\input{tex/tab_emo}

\vspace{-1em}
\subsection{Results}\vspace{-.5em}
Our evaluation focuses on two aspects: (i) the degree to which the target speaker identity is suppressed in the opt-out set, quantified by reductions in SIM, and (ii) the fidelity of the generated speech to the input text, measured by WER.
Compared to optimization-based unlearning methods (\ie, SGU~\cite{kim2025not} and TGU~\cite{kim2025not}), which require careful verification of both retain and opt-out unlearning, \ours applies steering only during synthesis for opt-out speakers, leaving generations for the retain speakers identical to those of the baseline model.

\tparagraph{Seen opt-out speakers.}
Table~\ref{tab:table1} reports the results on the seen opt-out set, demonstrating how well the original zero-shot TTS performance is preserved after unlearning on LibriSpeech.
The results show that \ours achieves a substantial reduction in SIM-SO, confirming the effective suppression of the target speaker’s identity. 
At the same time, WER-SO score on \ours outperforms SGU and TGU, which cost tremendous training, \eg, 48 and 430 hours for 2 A6000 GPUs, respectively.
This outcome—strong suppression of identity with minimal impact on content fidelity—provides direct evidence that \ours appropriately adjusts hidden identity activations rather than introducing indiscriminate perturbations.
While SGU shows the highest Spk-ZRF-SO and the lowest SIM-SO, SIM-R also declines substantially, indicating that it generates random voice for both retain and opt-out speakers. 
TGU fails to achieve sufficient identity suppression, as evidenced by the relatively modest reduction in SIM-SO.
Moreover, the concurrent decrease in SIM-R and increase in both WER-R and WER-SO indicate that TGU also suffers from degraded unlearning performance.

\tparagraph{Generalization to unseen opt-out speakers.} 
We further evaluate \ours on unseen opt-out speakers from LibriSpeech~\cite{panayotov2015librispeech}.
To facilitate the interpretation of such unseen speakers, \cref{tab:table2} and \cref{tab:emo} include zero-shot baseline performance of F5-TTS, which aims to preserve reference speaker identity, in contrast to our objective of suppressing opt-out identities.
As shown in \cref{tab:table2}, \ours improves both SIM-UO and Spk-ZRF-UO, despite a sligt degradation in WER-UO.
These results suggest that identity unlearning can generalize to unseen speakers without retraining.
It highlights the potential of inference-time, training-free unlearning as a practical direction for zero-shot TTS.

\tparagraph{Emotion preservation.} 
We measure SIM-UO and SIM-Emo using the emotional speech dataset~\cite{cremad} to verify whether emotion attributes are preserved even after applying \ours in a zero-shot setting.
To establish an upper bound for emotion preservation, we generate voices using F5-TTS conditioned on both speech and text prompts, which does not involve unlearning.
In \cref{tab:emo}, \ours yields a noticeably lower SIM-UO score than F5-TTS, confirming effective unlearning of opt-out speakers.
Concurrently, SIM-Emo remains largely preserved and is comparable to that of F5-TTS.
This indicates that \ours achieves effective unlearning (or prohibition) while maintaining non-speaker attributes beyond speaker identity.
\vspace{-.7em}
\input{tex/table_ablation_1}
\input{tex/table_ablation_2}

\vspace{-.7em}
\subsection{Ablation study}\vspace{-.5em}
\tparagraph{Layer filtering criteria.}
We examine how different layer filtering criteria affect unlearning by steering only layers where their similarity is below `$\mu-\sigma$', `$\mu$', `$\mu+\sigma$', or all layers. \cref{tab:layer_band} shows that the `$\mu+\sigma$' criterion provides the most balanced performance over all metrics.
While steering all layers yields marginal reductions in SIM-SO compared to ‘$\mu+\sigma$’, it incurs a disproportionately larger increase in WER-SO.
Additionally, stricter thresholds (`$\mu$' or `$\mu-\sigma$') yield weaker suppression for identity removal. 
These results indicate that `$\mu+\sigma$' effectively targets identity-specific signals while avoiding disruption of phonetic fidelity.

\tparagraph{The pool size of ID-prototype.}
\cref{tab:remain_num} examines the performance varying the retain-speaker pool size $N$=$\{10,30,50\}$.
Under the `$\mu+\sigma$' criterion, $N=30$ provides the best overall balance, achieving the lowest SIM and WER on seen data while remaining competitive on unseen data.
While $N$=$50$ performs best on unseen data, it degrades performance on seen data, motivating our choice of $N$=$30$.  
\vspace{-.7em}

%% file: tex/tab_libri.tex

\begin{table}[!t]
    \centering
    \small
    \resizebox{\linewidth}{!}{
    \begin{tabular}{lcccc}
        \toprule
        Methods & Unlearning & WER-UO $\downarrow$ & SIM-UO  ${\downarrow}$ & Spk-ZRF-UO $\uparrow$ \\
        \midrule
        F5-TTS~\cite{chen2024f5}    & \xmark       & 2.03 & 0.668 & 0.906 \\
        \ours    & \cmark            & 3.26 & 0.488 & 0.913\\
        \bottomrule
    \end{tabular}
    }
    \vspace{-.7em}
    \caption{Performance on unseen opt-out set (-UO) in LibriSpeech.}\label{tab:table2}\vspace{-.5em}
\end{table}

%% file: tex/tab_emo.tex
\begin{table}[!t]
    \centering
    \small
    \resizebox{0.8\linewidth}{!}{
    \begin{tabular}{lcccc}
        \toprule
        Methods & Unlearning & SIM-UO $\downarrow$ & SIM-Emo$\uparrow$ \\
        \midrule
        F5-TTS~\cite{chen2024f5} & \xmark & 0.217 & 0.732 \\
        \ours           & \cmark & 0.131 & 0.723 \\
        \bottomrule
    \end{tabular}
    }
    \vspace{-.5em}
    \caption{Evaluation for emotion preservation and unlearning performance on unseen opt-out set (-UO) in CREMA-D.}\label{tab:emo}
\end{table}

%% file: tex/table_ablation_1.tex
\begin{table}[t]
\centering
\resizebox{\linewidth}{!}{
\setlength{\tabcolsep}{2pt}
\begin{tabular}{lcccccc}
\toprule
$\tau$ & SIM-SO$\downarrow$ & WER-SO$\downarrow$ & Spk-ZFR-SO$\uparrow$ & SIM-UO$\downarrow$ & WER-UO$\downarrow$ & Spk-ZRF-UO$\uparrow$ \\
\midrule
$\mu-\sigma$ & 0.567 & 3.51 & 0.926 & 0.551 & 2.30 & 0.913 \\
$\mu$        & 0.538 & 3.35 & 0.926 & 0.494 & 2.81 & 0.913 \\
$\mu+\sigma$ & 0.477 & 3.25 & 0.929 & 0.488 & 3.26 & 0.913 \\
all          & 0.462 & 3.71 & 0.928 & 0.491 & 3.12 & 0.912 \\
\bottomrule
\end{tabular}
}\vspace{-.7em}
\caption{Performance over different layer selection strategies.}
\label{tab:layer_band}\vspace{-.5em}
\end{table}

%% file: tex/table_ablation_2.tex
\begin{table}[t]
\centering
\resizebox{\linewidth}{!}{
\setlength{\tabcolsep}{2pt}
\begin{tabular}{l ccc ccc}
\toprule
\# & SIM-SO$\downarrow$ & WER-SO$\downarrow$ & Spk-ZRF-SO$\uparrow$ & SIM-UO$\downarrow$ & WER-UO$\downarrow$ & Spk-ZRF-UO$\uparrow$ \\
\midrule
N=10 & 0.535 & 3.81 & 0.927 & 0.532 & 3.04 & 0.913 \\
N=30 & 0.477 & 3.25 & 0.929 & 0.488 & 3.26 & 0.913 \\
N=50 & 0.525 & 3.71 & 0.930 & 0.484 & 2.35 & 0.917 \\
\bottomrule
\end{tabular}
}\vspace{-.7em}
\caption{Performance over different numbers of retain speakers. 
}
\label{tab:remain_num}
\end{table}

%% file: tex/5_conclusion.tex
\section{Conclusion}\vspace{-.5em}
We introduce \ours, a training-free framework for opt-out speaker unlearning in zero-shot TTS for the first time.
\ours compares the hidden activations of the target speaker desired to unlearn against an averaged ID-prototype embedding, then steers the identity basis of activation.
We believe this paradigm shift offers a practical foundation for future generative speech systems, establishing a scalable solution to user-driven privacy requests.

\vspace{.1em}
\footnotesize{
\tparagraph{Acknowledgements.}
This work was supported by the National Research Foundation of Korea(NRF) grant
funded by the Korea government(MSIT) (RS-2025-16065706), Global - Learning \& Academic research institution for Master’s·PhD students, and Postdocs(G-LAMP) Program of the National Research Foundation of Korea(NRF) grant funded by the Ministry of Education(No. RS-2025-25442252), and the Ewha Womans University Research Grant of 2025.
}